\documentclass{appolb}
\usepackage{epsfig}

\begin{document}

\title{Spontaneous currents in a ferromagnet - normal metal - superconductor 
       trilayer
}

\author{M. Krawiec
\address{Institute of Physics and Nanotechnology Center, M. Curie-Sk\l odowska
         University, pl. M. Curie-Sk\l odowskiej 1, 20-031 Lublin, POLAND}
\and
J. F. Annett, B. L. Gy\"{o}rffy 
\address{H. H. Wills Physics Laboratory, University of Bristol, Tyndal Ave.,
         Bristol BS8 1TL, UK}
}
\maketitle


\begin{abstract}

We discuss the ground state properties of the system composed of a normal
metal sandwiched between ferromagnet and superconductor within a tight binding
Hubbard model. We have solved the spin-polarized Hartree-Fock-Gorkov equations
together with the Maxwell's equation (Ampere's law) and found a proximity 
induced Fulde-Ferrell-Larkin-Ovchinnikov (FFLO) state in this system. Here we 
show that the inclusion of the normal metal layer in between those subsystems 
does not necessarily lead to the suppression of the FFLO phase. Moreover, we 
have found that depending on the thickness of the normal metal slab the system 
can be switched periodically between the state with the spontaneous current 
flowing to that one with no current. All these effects can be explained in 
terms of the Andreev bound states formed in such structures.

\end{abstract}

  
\section{\label{intro} Introduction}

The proximity effect between a ferromagnet and a superconductor has attracted 
much attention recently due to its potential applications in such areas of 
technology as magnetoelectronics \cite{Bauer} or quantum computing 
\cite{Blatter}. The proximity effect in ferromagnet-superconductor (F-S) 
structures is also important from the point of view of the scientific interest 
as it allows the study of the interplay between ferromagnetism and 
superconductivity \cite{Berk}. 

Due to time reversal symmetry breaking the proximity effect in F-S structures 
leads to new phenomena, not observed in usual normal metal - superconductor 
(N-S) proximity systems. Those are: oscillations of the superconducting 
transition temperature \cite{Wong}, density of states \cite{Kontos} and 
superconducting pairing amplitude in F-S multilayers, Josephson $\pi$-junction 
behavior in S-F-S heterostructures \cite{Ryazanov}, a giant mutual proximity 
effect in S-F systems \cite{Petrashov}, spin valve \cite{Tagirov} or 
spontaneous currents in F-S bilayers \cite{MK_1,MK_2,BLG}. For review of the 
literature on those and related effects see \cite{Izyumov}. 

We have recently examined F-S proximity system and found that a spontaneously 
generated current flows on both sides of the interface \cite{MK_1,MK_2,BLG}.
Such a current, which flows in the ground state of the system is a hallmark of
the Fulde-Ferrell-Larkin-Ovchinnikov (FFLO) like state, originally predicted 
for a bulk superconductor in a magnetic exchange field acting on spins only 
\cite{FFLO}. It is the purpose of the present paper to see what the effect on 
the FFLO state will have a normal metal sandwiched between ferromagnet and 
superconductor (F-N-S structure). Will the spontaneous current still be 
generated  ? If so, how will it be modified ? In the rest of the paper we show 
that the presence of the normal metal has nontrivial consequences on the 
generation of the spontaneous currents and on the FFLO state in general.


\section{\label{model} The Model}

\begin{figure}[h]
\begin{center}
 \resizebox{0.7\linewidth}{!}{
  \includegraphics{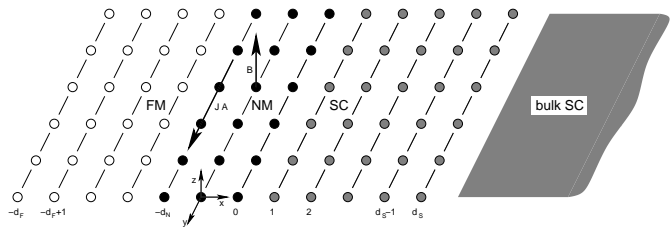}}
\end{center}
 \caption{\label{Fig1} Schematic view of our model system. The rows of empty 
          circles ($n = -d_F,...,-d_N$) describe ferromagnetic layers, filled
	  ($n = -d_N,...,0$) - normal region, while shaded ($n = 1,...$) -
	  superconducting one. Directions of the magnetic field ${\bf B}$,
	  vector potential ${\bf A}$ and current ${\bf J}$ are also indicated.}
\end{figure}

Our model system is schematically shown in Fig. \ref{Fig1}. 
It consists of a ferromagnet (layers $n = -d_F,...,-d_N$), a normal 
(paramagnetic) region (layers $n = -d_N,...,0$) and a semi-infinite 
superconductor (layers $n = 1,...$). 

The model Hamiltonian is given by single orbital Hubbard model
\begin{eqnarray}
H = \sum_{ij\sigma} \left[ t_{ij} + 
(\varepsilon_{i\sigma} - \mu) \delta_{ij} \right] c^+_{i\sigma} c_{j\sigma} +
\frac{1}{2} \sum_{i\sigma} U_i \hat n_{i\sigma} \hat n_{i -\sigma} ,
\label{Hamilt}
\end{eqnarray}
where, in the presence of a vector potential, the nearest neighbor hopping
integral is 
$t_{ij} = -t e^{-ie \int^{{\bf r}_j}_{{\bf r}_i} {\bf A}({\bf r}) 
\cdot d{\bf r}}$. The site energy levels $\varepsilon_{i\sigma}$ are equal to 
$\frac{1}{2} E_{ex} \sigma$ in the ferromagnet and $0$ in the normal and 
superconducting region, and $\mu$ is the chemical potential. $U_i$ is the 
on-site electron-electron interaction, which is assumed to be negative on 
superconducting side and zero elsewhere, $c^+_{i\sigma}$ ($c_{i\sigma}$) are 
the usual electron creation (annihilation) operators and 
$\hat n_{i\sigma} = c^+_{i\sigma} c_{i\sigma}$ is the electron number operator. 

We study the above model in the spin-polarized Hartree-Fock-Gorkov (SPHFG) 
approximation, assuming the Landau gauge for the magnetic field 
${\bf B} = (0, 0, B_z(x))$, thus ${\bf A} = (0, A_y(x), 0)$ (see Fig. 
\ref{Fig1}). In the following we assume periodicity of our model in the 
direction parallel to the interface and therefore we work in ${\bf k}$ space 
in the $y$ direction but in real space in the $x$ direction. 

As usual, self-consistency is assured by the relations determining SC order
parameter $\Delta_n$, current $J_{y \uparrow (\downarrow)}(n)$ and the vector
potential $A_y(n)$ on each layer $n$
\begin{eqnarray}
\Delta_n = U_n \chi_n = 
- \frac{U_n}{\pi} \sum_{k_y} \int d\omega {\rm Im} G^{12}_{nn}(\omega, k_y) 
f(\omega) ,
\label{Delta}
\end{eqnarray}
\begin{eqnarray}
J_{y\uparrow (\downarrow)}(n) = 
-\frac{2 e t}{\pi} \sum_{k_y} sin(k_y - e A_y(n)) \int d\omega {\rm Im} 
G^{11 (33)}_{nn}(\omega, k_y) f(\omega) ,
\label{current}
\end{eqnarray}
\begin{eqnarray}
A_y(n+1) - 2 A_y(n) + A_y(n-1) = -4 \pi J_y(n) ,
\label{vec_pot}
\end{eqnarray}
where $G^{\alpha\beta}_{nm}(\omega, k_y)$ is the $4 \times 4$ Nambu Green
function and $f(\omega)$ the Fermi distribution. Equation (\ref{vec_pot}) is a
lattice version of the Ampere's law 
$\frac{d^2 A_y(x)}{d x^2} = - 4 \pi J_y(x)$. The above equations have to be
solved self-consistently, using the method described in \cite{MK_1}. 


\section{\label{results} Results}

\begin{figure}[h]
\begin{center}
 \resizebox{\linewidth}{!}{
  \includegraphics{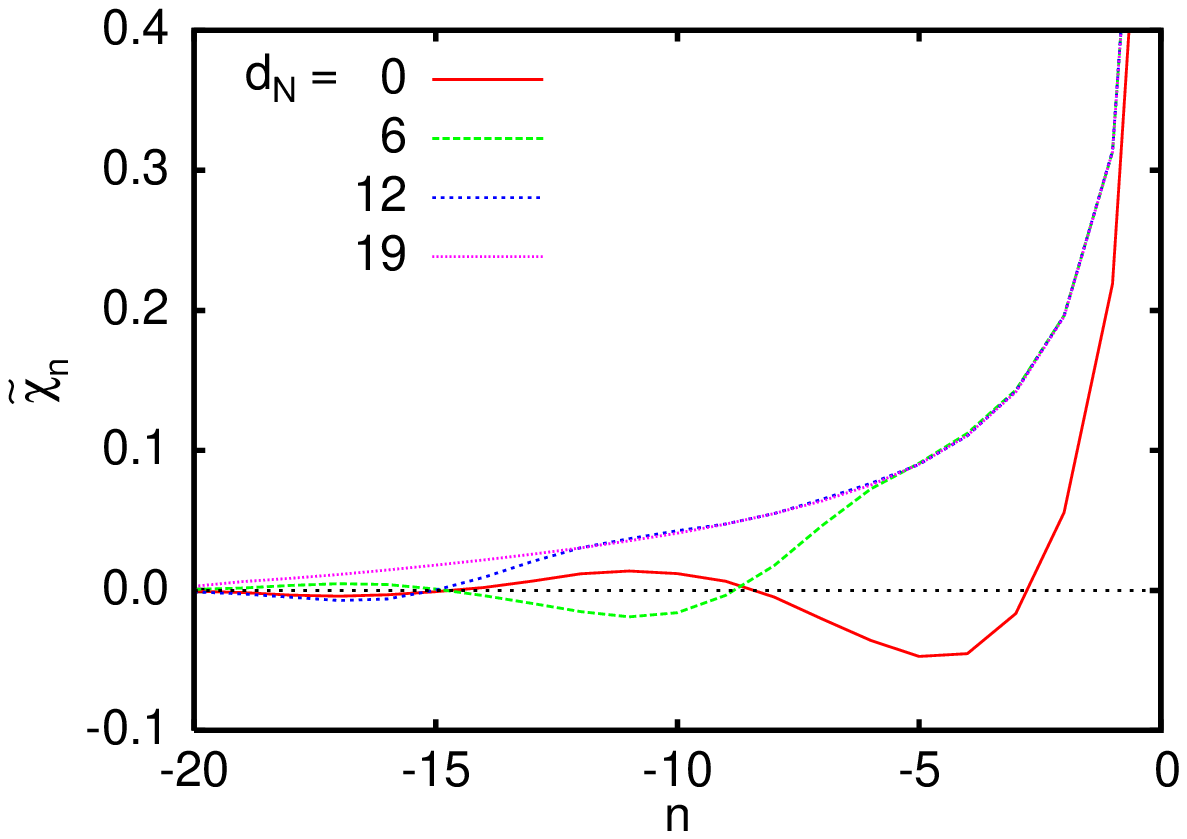}
  \includegraphics{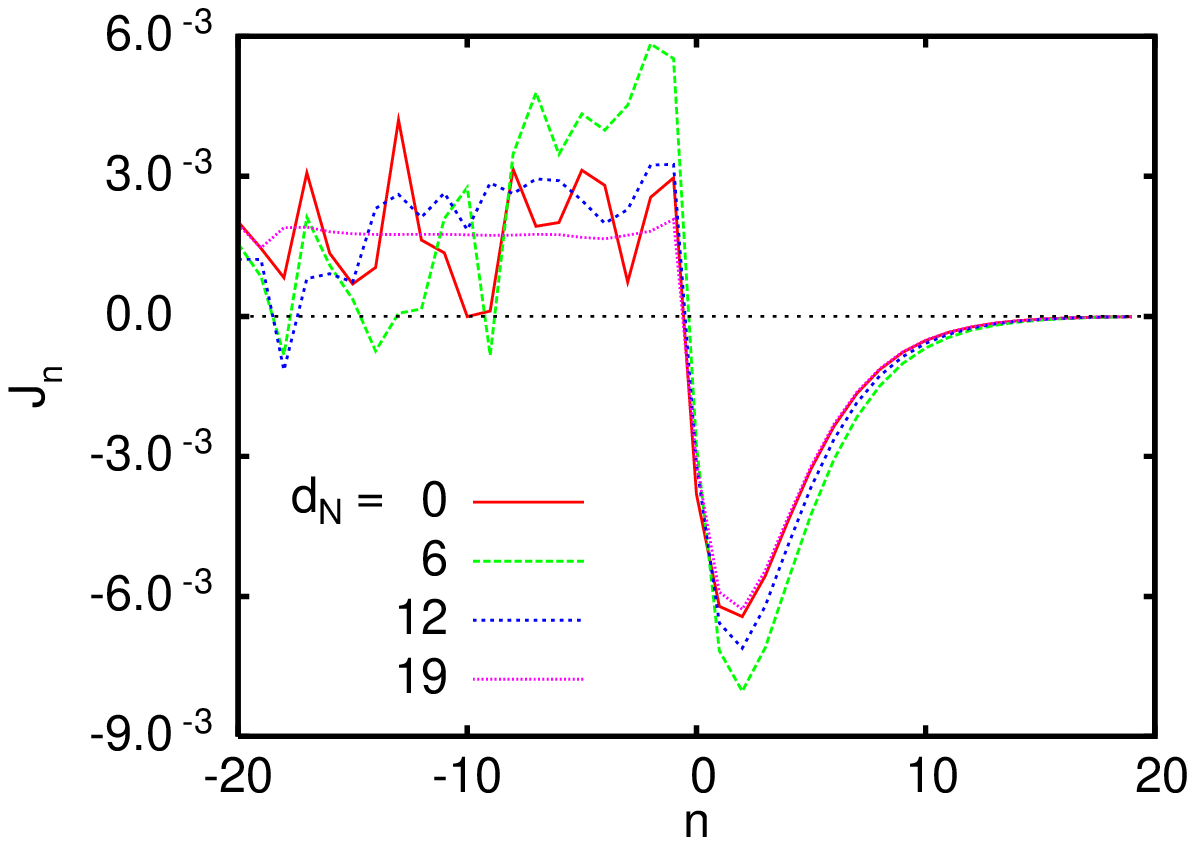}
  }
\end{center}
 \caption{\label{Fig2} Spatial dependence of the normalized superconducting 
          pairing amplitude $\tilde \chi_n = \chi_n/\chi_{bulk}$ (left panel) 
	  and spontaneous current $J_n$ (right panel) for a number of $d_F$ and 
	  $d_N$ ($d_F + d_N = 20$) layers. The model parameters are: 
	  $E_{ex} = 0.517$, $\Delta_S = 0.528$ and $T = 10^{-2}$ in units of 
	  $t$.}
\end{figure}

Figure \ref{Fig2} shows the spatial dependence of the normalized 
superconducting pairing amplitude $\tilde \chi_n = \chi_n/\chi_{bulk}$ (left 
panel) and spontaneous current $J_n$ (right panel) for a number of normal metal 
$d_N$ layers ($d_F = 20 - d_N$).
As in our previous work on  F-S structures \cite{Buzdin_1,MK_1,MK_2,BLG}, the 
superconducting pairing amplitude shows clear oscillations in the ferromagnet. 
The period of the oscillations is related to the ferromagnetic coherence length 
$\xi_F = 2t/E_{ex}$, i.e. $\chi_n \propto \frac{sin(n/\xi_F)}{(n/\xi_F)}$. No 
such effect is observed in the N region. One could expect that the inclusion of 
the normal metal suppresses the oscillations of $\chi_n$ in the F region. 
Interestingly, the amplitude of the $\chi_n$ oscillations on the ferromagnetic 
side remains almost unchanged. Only the phase of the oscillations can change. 
Moreover, as for F-S bilayer \cite{MK_1}, the vanishing of the pairing 
amplitude at $n = -d_F$ is related to the crossing of the Andreev bound states 
(ABS) through the Fermi energy of the system. Furthermore, in such a situation, 
spontaneous current is generated \cite{MK_1}. The current flowing produces a 
magnetic field, which splits this zero energy ABS, thus lowering the total 
energy of the system. The energy of this splitting is given in our model by 
$\delta \approx 2 e t \bar{A}_y$, where $\bar{A}_y$ is the layer averaged 
vector potential. 

As we already mentioned, the zero energy ABS is responsible for the generation 
of the spontaneous current. The typical distribution of such current, flowing
parallel to the interface, is shown in the right panel of Fig. \ref{Fig2}. The 
current flows in the negative $y$ direction on the superconducting side and in 
the positive direction in the whole $F-N$ region, giving a total current equal 
to zero, as should be in the true ground state. Interestingly, the current in 
the ferromagnet shows also oscillatory behavior, however, with a different 
period than the pairing amplitude does. The oscillations of the current are 
related to the oscillations of the density of states at the Fermi energy 
\cite{MK_1,MK_2}. On the other hand, in the N region, the distribution of the 
current is rather smooth, similarly to the pairing amplitude (compare the solid 
and thin dotted lines in Fig. \ref{Fig2}).

\begin{figure}[h]
\begin{center}
 \resizebox{0.5\linewidth}{!}{
  \includegraphics{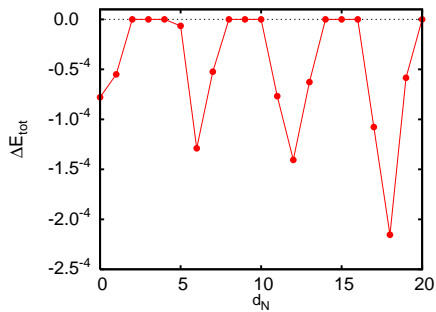}
  }
\end{center}
 \caption{\label{Fig3} The difference in total energy between solutions with 
          and without current.}
\end{figure}

The state with spontaneous current is a true ground state, as it has lower 
energy than the state with no current. This can be seen in Fig. \ref{Fig3}, 
which shows the change in total energy of the system $\Delta E_{tot}$ between 
the solution with spontaneous current and the one where the current is 
constrained to be zero is shown. 
Another important finding is that the system can be periodically switched 
between the states with and without spontaneous current by increasing of the 
thickness of the normal metal region. For $d_F + d_N = 20$ layers and 
$E_{ex} = 0.517$ the period is equal to six (see Fig. \ref{Fig3}). 

All above the results show that inclusion of the normal metal slab between the 
ferromagnet and superconductor has non-trivial effects on the physics of the
proximity induced FFLO state. It cannot be argued that the normal metal simply
acts as interface transparency, as in our case the N slab is in the clean 
(ballistic) regime, and thus does not suppress the proximity effect. However, 
it leads to the suppression of the oscillatory behavior of the pairing 
amplitude, spontaneous current and the density of states at the Fermi energy. 
Moreover, it strongly modifies the positions of the ABS, which leads to the 
periodical switching between the states with and without current. Interface 
transparency also leads to such periodical switching of the current  
\cite{MK_2}, however, at the same time it also kills the proximity effect, 
suppressing the current and changing the period of the pairing amplitude 
oscillations. No such effects are caused by the inclusion of the normal metal 
slab. Perhaps the effect of the N slab would be more similar to the effect of
reduced interface transparency if the normal metal was in the dirty regime.


\section{\label{conclusions} Conclusions}
In conclusion we have studied the ground state properties of F-N-S proximity
system. We have observed oscillatory behavior of the superconducting pairing
amplitude in ferromagnet, but not in the normal region. Similarly to F-S 
structures, we have found spontaneously generated currents flowing in the whole 
F and N regions and within a distance of a few $\xi_S$ on superconducting side. 
Interestingly, the system can be switched periodically (changing $d_N$) between 
the states with and without the spontaneous current. All this suggests possible 
realization of the proximity induced FFLO state in F-N-S heterostructure. 


\section*{Acknowledgements}
This work has been partially supported by the grant no. PBZ-MIN-008/P03/2003,
and by the ESF network programme AQD JJ.



\end{document}